\documentstyle[12pt]{article}

\newcommand{\be}{\begin{equation}}
\newcommand{\ee}{\end{equation}}
\newcommand{\Dlt}{\Delta}
\newcommand{\dlt}{\delta}
\newcommand{\om}{\omega}
\newcommand{\ep}{\varepsilon}

\newcommand{\br}{{\bf r}}
\newcommand{\bk}{{\bf k}}

\newcommand{\bt}{\beta}
\newcommand{\al}{\alpha}

\newcommand{\ra}{\rightarrow}
\newcommand{\sgm}{\sigma}

\newcommand{\prt}{\partial}
\newcommand{\cF}{{\cal F}}
\newcommand{\dgr}{\dagger}
\newcommand{\vp}{\varphi}

\begin{document}

\begin{center}
{\Large{\bf Nonequivalent operator representations for Bose-condensed 
systems} \\ [5mm]
V.I. Yukalov} \\ [3mm]

{\it Institut f\"ur Theoretische Physik, \\
Freie Universit\"at Berlin, Arnimallee 14, D-14195 Berlin, Germany \\ 
and \\
Bogolubov Laboratory of Theoretical Physics, \\
Joint Institute for Nuclear Research, Dubna 141980, Russia}

\end{center}

\vskip 1cm

\begin{abstract}

The necessity of accurately taking into account the existence of 
nonequivalent operator representations, associated with canonical 
transformations, is discussed. It is demonstrated that Bose systems in 
the presence of the Bose-Einstein condensate and without it correspond to 
different Fock spaces, orthogonal to each other. A composite representation 
for the field operators is constructed, allowing for a self-consistent 
description of Bose-condensed systems. Equations of motion are derived 
from the given Hamiltonian, which guarantees the validity of conservation
laws and thermodynamic self-consistency. At the same time, the particle
spectrum, obtained either from diagonalizing this Hamiltonian or from 
linearizing the field-operator equations of motion, has no gap. The condition 
of the condensate existence assures the absence of the gap in the spectrum, 
irrespectively to the approximation involved. The suggested self-consistent 
theory is both conserving and gapless.

\end{abstract}

\vskip 1cm


\vskip 1cm


\newpage

\section{Introduction}

Physics of Bose-condensed systems has received in recent years much 
consideration, both in experiment and theory. Numerous references on 
current literature can be found in the review articles [1--5]. The history 
of works on Bose-Einstein condensation is surveyed in Ref. [6].

The microscopic theory of Bose-condensed systems is based on the Bogolubov 
ideas, which have been applied to a low-temperature dilute gas in the 
pioneering papers [7,8] and expounded in detail in the books [9,10]. A 
key point in the Bogolubov approach is the separation of condensate on 
the level of the field operators, with replacing the condensate operator 
by a nonoperator quantity, playing the role of a condensate wave function. 
This procedure is asymptotically exact in the thermodynamic limit [11]. 
In general terms, this is formulated as the Bogolubov shift, when the 
field operator is shifted by the condensate wave function. This approach 
gives for the dilute Bose gas the famous Bogolubov spectrum having at 
low momenta the phonon character [7--10]. Beliaev [12,13] included in 
the Bogolubov approach interactions between quasiparticles, obtaining, 
by means of perturbation theory at zero temperature, attenuation 
parameters. Hugenholtz and Pines [14] formulated the condition on the 
chemical potential, providing a gapless quasiparticle spectrum at zero 
temperature. This condition was generalized to finite temperatures by 
Bogolubov [9,10] and Hohenberg and Martin [15]. The Beliaev zero-temperature 
calculations were extended to finite temperatures by Popov and Faddeev [16] 
and Popov [17--19]. The Bogolubov theory was applied to inhomogeneous systems 
by Fetter [20]. The Beliaev-Popov approach was extended to nonuniform trapped 
gases by Fedichev and Shlyapnikov [21] and by Morgan et al. [22--25].

In the lowest dilute-gas approximations, such as the Bogolubov and 
Beliaev-Popov approximations, the quasiparticle spectrum is gapless, in 
agreement with the Goldstone and Bogolubov theorems [9,10,15]. However, 
in higher approximations, the spectrum acquires a nonphysical gap, which 
happens, e.g., in the Hartree-Fock-Bogolubov approximation. The standard 
remedy for removing the gap is to construct higher-order vertex equations, 
such as the Bethe-Salpeter equation or Lipmann-Schwinger equation, including 
additional terms into the theory in order to compensate the gap [26]. But
so constructed compensations are not a result of a self-consistent scheme,
in which all equations would follow from one generally prescribed procedure. 
As a consequence, the resulting modifications suffer from violating 
conservations laws and thermodynamic consistency. In order to preserve 
conservation laws, all equations in the theory must be derivable from 
a given Hamiltonian or some variational functional [27,28]. For one 
chosen approximation, it is always possible to construct an artificial 
variational functional that would yield a gapless spectrum and even 
preserving conservation laws [29,30]. However, this cannot be considered 
as a self-consistent theory allowing for a systematic derivation of other
approximations. In this way, all theories of Bose-condensed systems suffer
either from the absence of conservation laws and thermodynamic consistency 
or from the presence of an unphysical gap in the quasiparticle spectrum. 
A detailed classification of theories onto either conserving but gapless 
or gapless but nonconserving has been done by Hohenberg and Martin [15].

The aim of the present paper is to suggest a self-consistent theory that 
would be both conserving as well as gapless. To be conserving, the theory
must be based on one microscopic Hamiltonian, from which all equations
would follow. To be gapless, the theory requires an accurate mathematical
formulation, taking account of the existence of the Bose-condensed fraction.
Such an accurate mathematical formulation can be done by explicitly taking 
into consideration the appearance of nonequivalent operator representations
associated with canonical commutation relations [31]. Since the theory to 
be developed rests on the notion of nonequivalent operator representations,
it is necessary, first of all, to specify its meaning for Bose-condensed
systems, which is done in Sections 2 and 3. Then, in Section 4, a composite
representation is defined, making it possible to accurately describe 
Bose-condensed systems. Equations of motion are derived in Section 5 from 
a microscopic Hamiltonian based on the composite representation. To make 
the resulting equations more transparent, they are illustrated for the 
case of a uniform dilute gas in Section 6. The generalization to nonuniform 
matter is given in Section 7. Everywhere the system of units is employed,
where the Planck constant $\hbar=1$ and the Boltzmann constant $k_B=1$ 
are set to unity. The material of this paper is based on the Lectures [32].

\section{Bogolubov shift}

The Bogolubov shift of the field operator is commonly defined as
\be
\label{1}
\psi(\br,t) =\eta(\br,t) + \psi_1(\br,t) \; ,
\ee
where $\psi(\br,t)$ and $\psi_1(\br,t)$ satisfy the Bose commutation 
relations and $\eta(\br,t)$ is a condensate wave function, assumed to be 
not an identical zero. In what follows, for the sake of brevity, the time 
dependence often will not be shown explicitly, so that we shall write 
$\psi(\br)$, $\eta(\br)$, and $\psi_1(\br)$, though keeping in mind that,
in general, these quantities do depend on time.

Field operators act on functions pertaining to a Fock space. Let the field
operators $\psi(\br)$ and $\psi^\dgr(\br)$ be defined on the corresponding
Fock space $\cF(\psi)$. Let $|0>$ be the vacuum state of $\cF(\psi)$, such 
that for any $\br$ one has
\be
\label{2}
\psi(\br)|0>\;  = \; 0 \; .
\ee
Then any function $\vp\in\cF(\psi)$ can be generated by a repeated action of
$\psi^\dgr(\br)$ according to the rule
$$
\vp = \sum_{n=0}^\infty\; \frac{1}{\sqrt{n!}} \;
\int f_n(\br_1,\ldots,\br_n)\prod_{i=1}^n \psi^\dgr(\br_i)\; 
d\br_i|0> \; .
$$
In that sense, one says that the Fock space $\cF(\psi)$ is generated by 
$\psi^\dgr(\br)$.

It is easy to notice that the state $|0>$ is not a vacuum for $\psi_1(\br)$,
since
\be
\label{3}
\psi_1(\br)|0>\; = \; -\eta(\br)|0>\; \neq \; 0 \; .
\ee
A vacuum for $\psi_1(\br)$ is another state, which will be denoted as 
$|0>_1$, and which satisfies the vacuum-state equation
\be
\label{4}
\psi_1(\br)|0>_1 = 0 \; .
\ee
In line with the general rules [33], the creation operator $\psi_1^\dgr(\br)$ 
generates the Fock space $\cF(\psi_1)$. Clearly, the vacuum state $|0>_1$ is
not a vacuum for $\psi(\br)$, as far as
\be
\label{5}
\psi(\br)|0>_1 = \eta(\br)|0>_1 \neq 0 \; .
\ee
Thus, transformation (1) cannot be understood as an operator equality defined 
on one Fock space.

Let us introduce the operator
\be
\label{6}
\hat C \equiv \exp\left\{ \int \left [
\eta^*(\br)\psi(\br) - \eta(\br)\psi^\dgr(\br)\right ] \; d\br
\right \}
\ee
and its inverse
\be
\label{7}
\hat C^{-1} =  \exp\left\{ - \int \left [
\eta^*(\br)\psi(\br) - \eta(\br)\psi^\dgr(\br)\right ] \; d\br
\right \} \; .
\ee
With the help of these operators, relation (1) takes the form
\be
\label{8}
\psi(\br) = \hat C\psi_1(\br)\hat C^{-1}
\ee
and, conversely,
\be
\label{9}
\psi_1(\br) = \hat C^{-1}\psi(\br)\hat C \; .
\ee
By this definition, the operator $\hat C$ transforms functions from 
$\cF(\psi_1)$ into those in $\cF(\psi)$, while $\hat C^{-1}$ transforms 
the states of $\cF(\psi)$ into those of $\cF(\psi_1)$. In particular, the
relation between the vacua in $\cF(\psi)$ and in $\cF(\psi_1)$ is
\be
\label{10}
|0>_1 = \hat C^{-1}|0> \; ,
\ee
which is obvious from Eq. (9).

To better understand the connection between the spaces $\cF(\psi)$ and 
$\cF(\psi_1)$, we can use the Baker-Hausdorff formula for any two operators 
$\hat A$ and $\hat B$, whose commutator $[\hat A,\; \hat B]$ is proportional
to a unity operator. Then
$$
e^{\hat A +\hat B} = e^{\hat A} e^{\hat B}\exp\left ( -\; 
\frac{1}{2}\; [\hat A,\; \hat B] \right ) \; .
$$
Using this formula, operator (7) can be represented as
\be
\label{11}
\hat C^{-1} = \exp\left \{ \int \eta(\br)\psi^\dgr(\br)\; d\br\right \}
\exp\left \{ - \int \eta^*(\br)\psi(\br)\; d\br \right \} \; 
\exp\left\{ - \; \frac{1}{2}\int |\eta(\br)|^2\; d\br\right \} \; .
\ee
Form here, we have
\be
\label{12}
\hat C^{-1}|0>\; = \exp\left\{ -\; \frac{1}{2} \int |\eta(\br)|^2
d\br\right\} \exp \left \{ \int \eta(\br)\psi^\dgr(\br)\; d\br
\right \}|0>\; .
\ee
The right-hand side of Eq. (12) is nothing but the coherent state
$$
|\eta>\; = \eta_0\exp\left\{ \int \eta(\br)\psi^\dgr(\br)\; d\br
\right \}|0> \; ,
$$
normalized to unity, so that
$$
|\eta_0| =\exp\left\{ - \; \frac{1}{2} \; \int |\eta(\br)|^2 d\br
\right \} \; .
$$
By definition, the coherent state is given by the eigenproblem
\be
\label{13}
\psi(\br)|\eta>\; = \; \eta(\br)|\eta> \; ,
\ee
whose eigenvalue is a coherent field $\eta(\br)$. Hence, vacuum (10) is the
coherent state,
\be
\label{14}
|0>_1 \; = \hat C^{-1}|0>\; = |\eta>\; ,
\ee
and the condensate wave function in the Bogolubov shift (1) is nothing but the
coherent field
\be
\label{15}
\eta(\br) = \; <\eta|\psi(\br)|\eta> \; .
\ee
The scalar product of the vacua $|0>$ and $|0>_1$ is
\be
\label{16}
<0|0>_1 \; = \; <0|\eta>\; = \; \exp\left\{ -\;
\frac{1}{2}\; \int |\eta(\br)|^2\; d\br\right \} \; ,
\ee
as far as
$$
<0|\exp \left\{\int \eta(\br)\psi^\dgr(\br)\; d\br\right \}|0>\; = 
\; 1\; .
$$
The condensate wave function defines the condensate 
density
\be
\label{17}
\rho_0(\br) =|\eta(\br)|^2 \; .
\ee
Respectively, the latter determines the number of condensed particles
\be
\label{18}
N_0  =\int \rho_0(\br)\; d\br \; .
\ee
So, the scalar product (16) gives
\be
\label{19}
<0|\eta>\; = \; \exp\left ( - \; \frac{1}{2}\; N_0 \right ) \; .
\ee
Bose-Einstein condensation implies that the number of condensed particles 
$N_0$ is large, being of the order of the total number of particles $N$. 
Therefore product (19) is asymptotically zero,
$$
<0|\eta>\; \simeq \; 0  \qquad (N_0 \gg 1) \; .
$$
That is, the vacua $|0>$ and $|0>_1$ are asymptotically orthogonal. 
Similarly, all states generated by $\psi^\dgr(\br)$ from $|0>$ are 
orthogonal to the states generated by $\psi_1^\dgr(\br)$ from $|0>_1$. 
This is equivalent to saying that the Fock spaces $\cF(\psi)$ and 
$\cF(\psi_1)$ are asymptotically orthogonal to each other. There is 
the sole state that is shared by both these spaces. This is the coherent 
state $|\eta>$ of $\cF(\psi)$, which is at the same time the vacuum 
$|0>_1=|\eta>$ of $\cF(\psi_1)$. But since the spaces $\cF(\psi)$ and 
$\cF(\psi_1)$ are of continuous power, the sole intersection is called 
to be of zero measure.

The operator transformation (6), from $\cF(\psi_1)$ to $\cF(\psi)$, is not 
unitary. The inverse transformation (7), from $\cF(\psi)$ to $\cF(\psi_1)$, 
is not the same as $\hat C^+$, since it is not defined in one space. To be 
unitary, an operator and its Hermitian conjugate must be defined in one 
space. Therefore the Bogolubov shift (1) realizes the unitary nonequivalent 
representations of the field operators satisfying the Bose commutation 
relations. Strictly speaking, Eq. (1) has no sense of an operator equation, 
since its left-hand and right-hand sides are defined on different Fock 
spaces orthogonal to each other. More correctly, it has to be interpreted 
not as an equation but as a replacement
\be
\label{20}
\psi(\br) \longrightarrow \eta(\br) + \psi_1(\br) \; ,
\ee
with its left-hand side being defined on $\cF(\psi)$ and the right-hand side, 
on $\cF(\psi_1)$. 

As a physical interpretation, we may say that the field operator 
$\psi(\br)$ describes the Bose system {\it before} Bose-Einstein condensation 
has occurred, while the operator $\eta(\br)+\psi_1(\br)$ corresponds to that 
Bose system {\it after} the occurrence of Bose-Einstein condensation. This 
is somewhat analogous to the Van Hove picture of scattering [34,35], according
to which the particle states, before the scattering process has occurred, are 
orthogonal to their states after the scattering. That is, the Fock space of 
particles before the scattering is orthogonal to their Fock space after the 
scattering. Thence the particle field operators before and after the 
scattering process pertain to unitary nonequivalent representations of 
the same commutation relations.

It is important to emphasize that the right-hand side of replacement (20) 
clearly breaks the gauge symmetry. The latter could also be broken in the 
space $\cF(\psi)$ by adding to the Hamiltonian infinitesimal sources 
breaking this symmetry. However, irrespectively to whether the gauge symmetry 
in $\cF(\psi)$ is broken or not, this space remains always orthogonal to 
$\cF(\psi_1)$. This means that by breaking in $\cF(\psi)$ the gauge symmetry 
solely by infinitesimal sources (without the Bogolubov shift) one cannot make 
the transformation from $\cF(\psi)$ to $\cF(\psi_1)$. For the correct 
description of a Bose-condensed system it is not sufficient just to add to 
a Hamiltonian infinitesimal sources breaking the gauge symmetry, but it is 
necessary to make the Bogolubov shift, transferring the consideration to 
another space. Recall that separating in the Hamiltonian the condensate 
field operators is equivalent to the Bogolubov shift, because the condensate 
operators, under the thermodynamic limit, become nonoperator quantities 
representing the condensate wave function [9--11].

\section{Canonical transformations}

The Bogolubov shift (1) or (20) is an example of a simple canonical 
transformation realizing unitary nonequivalent operator representations. 
Another canonical transformation, constantly employed in the theory of Bose 
systems, is the Bogolubov canonical transformation
\be
\label{21}
a_k = u_k b_k + v_{-k}^* b_{-k}^\dgr
\ee
from the operators $a_k$ and $a_k^\dgr$ to $b_k$ and $b_k^\dgr$, where the 
label $k$ implies a set of quantum numbers, e.g., the momentum. The inverse 
transformation is 
\be
\label{22}
b_k = u_k^* a_k - v_k^* a_{-k}^\dgr \; .
\ee
In order that $a_k$ and $b_k$ would satisfy the same Bose commutation 
relations, the coefficient functions have to obey the normalization condition
$$
|u_k|^2 - |v_k|^2 =  1\; .
$$
Transformations (21) and (22) also realize unitary nonequivalent operator 
representations. Showing this below, we shall explain how this case differs 
from that considered in the previous section.

Let $|0>_a$ be the vacuum state for the operator $a_k$, so that
\be
\label{23}
a_k|0>_a = 0 \; .
\ee
Using the standard procedure [33], one can construct the Fock space $\cF(a_k)$ 
generated by $a_k^\dgr$. The state $|0>_a$ is not a vacuum for $b_k$, since
$$
b_k|0>_a = - v_k^* a_{-k}^\dgr|0>_a \; \neq 0 \; .
$$
The operators $b_k$ possess another vacuum $|0>_b$, for which 
\be
\label{24}
b_k|0>_b\; = 0 \; .
\ee
And $|0>_b$ is not a vacuum for $a_k$, because
$$
a_k|0>_b \; = v_{-k}^* b_{-k}^\dgr|0>_b \; \neq 0 \; .
$$
The Fock space $\cF(b_k)$ is generated by $b_k^\dgr$.

Usually, the coefficient functions can be chosen to be real and invariant 
under the change of $k$ onto $-k$, which we shall assume below setting
$$
u_k^* = u_{-k} = u_k \; , \qquad v_k^* = v_{-k} = v_k \; .
$$
Introducing the quantity
\be
\label{25}
\al_k \equiv \ln ( u_k + v_k) \; ,
\ee
one may write
\be
\label{26}
u_k ={\rm cosh}\al_k \; , \qquad v_k ={\rm sinh}\al_k \; .
\ee

Define the operator
\be
\label{27}
\hat B \equiv \exp\left \{ \frac{1}{2} \; \sum_k \al_k\left (
a_k a_{-k} - a_{-k}^\dgr a_k^\dgr\right ) \right \}
\ee
and its inverse
\be
\label{28}
\hat B^{-1} = \exp\left \{ -\;  \frac{1}{2} \; \sum_k \al_k\left (
a_k a_{-k} - a_{-k}^\dgr a_k^\dgr\right ) \right \} \; .
\ee
Using this, transformations (21) and (22) can be represented as
\be
\label{29}
a_k = \hat B b_k\hat B^{-1}
\ee
and, respectively, as
\be
\label{30}
b_k = \hat B^{-1} a_k \hat B \; .
\ee
The operator $\hat B$ transforms $\cF(b_k)$ into $\cF(a_k)$, while 
$\hat B^{-1}$ transforms $\cF(a_k)$ into $\cF(b_k)$. Thus, the vacua in 
these spaces are related by the transformation
\be
\label{31}
|0>_b\; = \hat B^{-1}|0>_a  \; .
\ee

Involving the Baker-Hausdorff formula, we can represent operator (28) as
\be
\label{32}
\hat B^{-1} = \exp\left ( \frac{1}{2} \; \sum_k \al_k a_{-k}^\dgr 
a_k^\dgr\right )
\exp\left ( - \; \frac{1}{2} \; \sum_k \al_k a_k a_{-k} \right )
\exp\left \{ -\; \frac{1}{4} \; \sum_k \al_k^2 (1+2 a_k^\dgr a_k) 
\right \} \; ,
\ee
keeping in mind its action on vacuum $|0>_a$, for which
$$
\sum_{kp} \al_k \al_p \left [ a_{-k}^\dgr a_k^\dgr, a_p a_{-p}\right ]
|0>_a \; = \sum_k \al_k^2 \left ( 1 + 2a_k^\dgr a_k\right )|0>_a\; =
\sum_k \al_k^2|0>_a \; .
$$
Then for vacuum (31), we find
\be
\label{33}
|0>_b \; = \hat B^{-1}|0>_a \; =\exp\left ( -\;  \frac{1}{4} \; \sum_k 
\al_k^2 \right ) \exp\left ( \frac{1}{2} \; \sum_k \al_k a_{-k}^\dgr 
a_k^\dgr\right )|0>_a \; .
\ee
The scalar product of the vacua $|0>_a$ and $|0>_b$ is
\be
\label{34}
_a<0|0>_b \; = \; \exp \left \{ -\; \frac{V}{4}\; \int
\al_k^2\; \frac{d\bk}{(2\pi)^3} \right \} \; ,
\ee
where $V$ is the system volume, and the equality
$$
_a<0|\exp\left ( \frac{1}{2} \;
\sum_k a^\dgr_{-k} a^\dgr_k\right )|0>_a \; = \; 1
$$ 
is taken into account. The integral in Eq. (34) is positive. As is evident, 
the vacua are asymptotically orthogonal, so that
$$
_a<0|0>_b\; \simeq 0  \qquad (V\ra \infty) \; .
$$
The Fock spaces $\cF(a_k)$ and $\cF(b_k)$ are mutually orthogonal in the 
thermodynamic limit. The operator $\hat B$ is not unitary. Thus, the 
canonical transformations (21) and (22) realize unitary nonequivalent
operator representations. Since the left-hand and right-hand sides of 
Eq. (21) are defined on different Fock spaces, it should be understood 
not as a straightforward operator equality, but as a replacement
\be
\label{35}
a_k \longrightarrow u_kb_k + v_{-k}^* b_{-k}^\dgr \; .
\ee

The main physical difference between the transformations (20) and (35) is 
that the latter is not related to the existence or absence of Bose-Einstein
condensate, while the Bogolubov shift (2) transforms a system without 
Bose condensate to a Bose-condensed system.

\section{Composite representation}

The description of a Bose system without Bose-Einstein condensate can be 
done in terms of one field operator $\psi(\br)$ defined on the Fock space 
$\cF(\psi)$ generated by $\psi^\dgr(\br)$ from a vacuum state $|0>$. But
a Bose-condensed system requires to employ two variables, the condensate 
wave function $\eta(\br)$ and the operator of noncondensed particles
$\psi_1(\br)$ defined on the Fock space $\cF(\psi_1)$ generated by 
$\psi_1^\dgr(\br)$ from the vacuum $|0>_1$. In order to stress that there
are two variables for a Bose-condensed system, it is possible to treat both 
condensed and noncondensed particles on an equal footing by introducing a
composite representation for the field operators.

As is shown above, the condensate corresponds to the coherent state. We 
may define the one-dimensional space
\be
\label{36}
\cF_0 \equiv \{ |\eta>\}
\ee
containing the coherent state for a field operator $\psi_0(\br)$, for 
which
\be
\label{37}
\psi_0(\br)|\eta>\; = \; \eta(\br)|\eta> \; .
\ee
The operator of noncondensed particles $\psi_1(\br)$ is defined on the Fock 
space $\cF(\psi_1)$ generated by $\psi_1^\dgr(\br)$ from the vacuum $|0>_1$. 
In order to avoid the double counting of the degrees of freedom, the
orthogonality condition
\be
\label{38}
\int \eta^*(\br)\psi_1(\br)\; d\br = 0
\ee
is imposed.

Since in the space $\cF(\psi_1)$ the gauge symmetry is explicitly broken, 
the statistical average $<\psi_1(\br)>_{\cF(\psi_1)}$ over this space may 
be not zero. This, however, would result in the nonconservation of quantum 
numbers. For instance, in the case of particles with nonzero spin, the latter 
would not be conserved. Or, Fourier transforming $\psi_1(\br)$ to $a_k$, and
having a nonzero statistical average $<a_k>_{\cF(a_k)}$, one would confront
the absence of momentum conservation. To eliminate such unpleasant features,
one defines the restricted space $\cF_1\subset\cF(\psi_1)$, such that the
restricted average
\be
\label{39}
<\psi_1(\br)>_{\cF_1}\; = 0
\ee
be zero. Formally, the restricted space is defined as 
\be
\label{40}
\cF_1 \equiv \left \{ \vp\in\cF(\psi_1)|\; <\psi_1(\br)>_{\cF_1}\; =0
\right \} \; .
\ee
An explicit and mathematically correct definition of the restricted space 
(40) can be done in terms of the weighted Hilbert spaces [36,37].

Let us now introduce the {\it composite Fock space}
\be
\label{41}
\tilde\cF \equiv \cF_0 \otimes \cF_1 \; ,
\ee
being a tensor product of the spaces (36) and (40). And let us define the
{\it composite field operator}
\be
\label{42}
\tilde\psi(\br) \equiv \psi_0(\br)\oplus \psi_1(\br) 
\ee
acting on the composite space (41). In what follows, we shall simplify the
notation often writing, instead of the sign of the direct summation $\oplus$,
just the usual signs of summation or subtraction. We shall also omit, as is 
usually done, the unity operators $\hat 1_0$ and $\hat 1_1$ defined in the
corresponding spaces $\cF_0$ and $\cF_1$.

The condensate wave function, according to definition (37), is the coherent
field
\be
\label{43}
\eta(\br) = \; <\eta|\psi_0(\br)|\eta>\; .
\ee
The number-of-particle operator of condensed particles, defined in the space
$\cF_0$, is
\be
\label{44}
\hat N_0 \equiv \int \psi_0^\dgr(\br)\psi_0(\br)\; d\br \; .
\ee
The condensate wave function is normalized to the number of condensed 
particles
\be
\label{45}
N_0 \equiv \; < \eta|\hat N_0|\eta>\; = \; 
\int |\eta(\br)|^2\; d\br \; .
\ee

The vectors $f\in\tilde\cF$, pertaining to the composite space (41), have
the structure of the tensor product
$$
f =|\eta> \otimes\; \vp \; ,
$$
in which $|\eta>\in\cF_0$ and $\vp\in\cF_1$. For any two vectors 
$f_m,f_n\in\tilde\cF$, the matrix element of the composite field operator
(42) is
\be
\label{46}
f_m^+\;\tilde\psi(\br) f_n = \vp_m^+ <\eta|\tilde\psi(\br)|\eta>
\vp_n =  \vp_m^+ \left [ \eta(\br) +\psi_1(\br)\right ]\vp_n \; .
\ee
Respectively, the matrix element of an operator $\hat A[\tilde\psi]$ 
from the algebra of local observables on $\tilde\cF$ takes the form
\be
\label{47}
f_m^+\;\hat A[\tilde\psi] f_n = \vp_m^+ <\eta|\hat A[\tilde\psi]|\eta>
\vp_n = \vp_m^+ \; \hat A[\eta+\psi_1]\vp_n \; .
\ee
The trace of such an operator is
$$
{\rm Tr}_{\tilde\cF} \hat A[\tilde\psi] = {\rm Tr}_{\cF_1}
<\eta|\hat A[\tilde\psi]|\eta>\; = \; {\rm Tr}_{\cF_1} 
\hat A[\eta+\psi_1] \; .
$$
The statistical average over the space $\tilde\cF$, with respect to a 
statistical operator $\hat\rho[\tilde\psi]$, taken at the initial time 
$t=0$, writes as
\be
\label{48}
<\hat A[\tilde\psi]>_{\tilde\cF} \; \equiv {\rm Tr}_{\tilde\cF}
\hat\rho[\tilde\psi] \hat A[\tilde\psi] = {\rm Tr}_{\cF_1}
\hat\rho[\eta+\psi_1]\hat A[\eta+\psi_1] \; .
\ee
Defining the average over $\cF_1$ as
\be
\label{49}
<\hat A[\eta+\psi_1]>_{\cF_1} \; \equiv {\rm Tr}_{\cF_1}
\hat\rho[\eta +\psi_1] \hat A[\eta +\psi_1] \; ,
\ee
we get
\be
\label{50}
<\hat A[\tilde\psi]>_{\tilde\cF}\; = \; <\hat A[\eta+\psi_1]>_{\cF_1} \; .
\ee

The operator of the total number of particles
\be
\label{51}
\hat N \equiv \int \tilde\psi^\dgr(\br) \tilde\psi(\br)\; d\br \; ,
\ee
with the composite field operator (42), keeping in mind the orthogonality 
condition (38), can be represented as the direct sum
\be
\label{52}
\hat N = \hat N_0 \oplus \hat N_1
\ee
of the condensed-particle operator (44) and the operator for the number 
of noncondensed particles
\be
\label{53}
\hat N_1 \equiv \int \psi_1^\dgr(\br) \psi_1(\br)\; d\br \; .
\ee
The average of the total number-of-particle operator (51), according to 
Eq. (50), becomes
\be
\label{54}
<\hat N>_{\tilde\cF}\; =  N_0 + <\hat N_1>_{\cF_1} \; .
\ee
The equality
\be
\label{55}
N = \; <\hat N>_{\tilde\cF}
\ee
for the total number of particles serves as a normalization condition, 
additional to the normalization (45) for the number of condensed particles.
Both these normalization conditions, (45) and (55), must be taken into 
account in a self-consistent theory.

\section{Equations of motion}

The energy operator is given by the Hamiltonian 
\be
\label{56}
\hat H = \int \tilde\psi^\dgr(\br) \left ( -\;
\frac{\nabla^2}{2m} + U \right ) \tilde\psi(\br) \; d\br + \frac{1}{2}
\int \tilde\psi^\dgr(\br)\tilde\psi^\dgr(\br')\Phi(\br-\br')
\tilde\psi(\br')\tilde\psi(\br)\; d\br d\br' \; ,
\ee
in which, $m$ is particle mass, $U=U(\br,t)$ is an external field, and
$\Phi(\br)=\Phi(-\br)$ is an interaction potential, assumed to be integrable, 
so that
$$
\left | \int_V \Phi(\br)\; d\br \right | < \infty \; .
$$
The field operators $\tilde\psi(\br)=\tilde\psi(\br,t)$ are given in the
Heisenberg representation, but the temporal variable $t$ is not explicitly 
shown just for making the formulas less cumbersome.

To take into consideration two normalization conditions, (45) and (55), 
it is necessary to introduce two Lagrange multipliers, $\mu_0$ and $\mu$, 
entering the grand Hamiltonian
\be
\label{57}
H[\tilde\psi] = \hat H - \mu_0 \hat N_0 - \mu \hat N \; .
\ee
Here $\mu$ is the conventional chemical potential and $\mu_0$ is an 
additional Lagrange multiplier guaranteeing the validity of the 
normalization condition (45).

With the composite field operator (42), the Hamiltonian (57) can be 
represented as a sum
\be
\label{58}
H[\tilde\psi] = \sum_{n=0}^ 4 H^{(n)} \; ,
\ee
in which the terms are classified according to the number of the 
noncondensed-particle operators in the products of each term. Thus, the
zero-order term contains only the condensate field operators,
$$
H^{(0)} = \int \psi_0^\dgr(\br) \left ( -\; \frac{\nabla^2}{2m} + 
U  - \mu_0 - \mu\right )\psi_0(\br)\; d\br + 
$$
\be
\label{59}
+ \frac{1}{2} \int 
\psi_0^\dgr(\br)\psi_0^\dgr(\br')\Phi(\br-\br')\psi_0(\br')\psi_0(\br)\;
d\br d\br' \; .
\ee
The first-order term is
$$
H^{(1)} = \int \psi_1^\dgr(\br)\left ( -\; \frac{\nabla^2}{2m} + U 
- \mu\right )\psi_0(\br)\; d\br + \int \psi_1(\br) \left ( 
-\; \frac{\nabla^2}{2m} + U - \mu\right ) \psi_0^\dgr(\br)\; d\br \; +
$$
\be
\label{60}
+ \int \Phi(\br-\br') \left [ \psi_1^\dgr(\br)\psi_0^\dgr(\br')
\psi_0(\br')\psi_0(\br) + \psi_0^\dgr(\br)\psi_0^\dgr(\br')\psi_0(\br')
\psi_1(\br)\right ]\; d\br d\br' \; .
\ee
However, a Hamiltonian term of the first order  in $\psi_1(\br)$ or
$\psi_1^\dgr(\br)$, being considered on the space $\cF_1$ defined in Eq. 
(40), is effectively zero due to the conservation condition (39). If a
Hamiltonian would contain such linear in $\psi_1(\br)$ or $\psi_1^\dgr(\br)$
terms, condition (39) could not be satisfied. The same can be said in another 
manner. To take into account restriction (39), we could add to the grand
Hamiltonian (57) one more term with a Lagrange multiplier assuring the 
validity of Eq. (39). To this end, this additional term has to be chosen 
so that to cancel the linear term (60). In any way,
\be
\label{61}
H^{(1)} = 0  \; .
\ee
For the quadratic term in Eq. (58), we have
$$
H^{(2)} = \int \psi_1^\dgr(\br) \left ( -\; \frac{\nabla^2}{2m} + U -
\mu\right ) \psi_1(\br)\; d\br \; + 
$$
$$
+ \int \Phi(\br-\br') \left [
\psi_0^\dgr(\br) \psi_1^\dgr(\br')\psi_1(\br')\psi_0(\br) +
\psi_0^\dgr(\br)\psi_1^\dgr(\br')\psi_0(\br')\psi_1(\br) + 
\right.
$$
\be
\label{62}
\left. + 
\frac{1}{2}\; \psi_0^\dgr(\br)\psi_0^\dgr(\br')\psi_1(\br')\psi_1(\br) +
\frac{1}{2}\; 
\psi_1^\dgr(\br)\psi_1^\dgr(\br')\psi_0(\br')\psi_0(\br)\right ] \;
d\br d\br' \; .
\ee
The third-order term is
\be
\label{63}
H^{(3)} = \int \Phi(\br-\br')\left [ 
\psi_0^\dgr(\br)\psi_1^\dgr(\br')\psi_1(\br')\psi_1(\br) +
\psi_1^\dgr(\br)\psi_1^\dgr(\br')\psi_1(\br')\psi_0(\br) \right ]\;
d\br d\br' \; .
\ee
In expressions (62) and (63), the symmetry $\Phi(\br)=\Phi(-\br)$ is used.
The last term is
\be
\label{64}
H^{(4)} = \frac{1}{2} \int 
\psi_1^\dgr(\br)\psi_1^\dgr(\br')\Phi(\br-\br') \psi_1(\br')\psi_1(\br)\;
d\br d\br' \; .
\ee

In order to preserve all conservation laws, such as the continuity equation, 
the density of current, and the energy-momentum conservation laws, all 
equations of motion must be derived from the same Hamiltonian. A theory
with such Hamiltonian-derivable equations will automatically be conserving.
Since we have two field-operator variables, we should have two types of 
equations of motion.

The first equation prescribes the evolution of the condensate operator
\be
\label{65}
i\; \frac{\prt}{\prt t}\; \psi_0(\br,t) = 
\frac{\dlt H[\psi_0\oplus\psi_1]}{\dlt\psi_0^\dgr(\br,t)} \; .
\ee
For the Hamiltonian (58), one has
$$
\frac{\dlt H[\psi_0\oplus\psi_1]}{\dlt\psi_0^\dgr(\br,t)} =
\frac{\dlt}{\dlt\psi_0^\dgr(\br,t)} \; \left ( H^{(0)} + H^{(2)} +
H^{(3)} \right ) \; .
$$
In the expressions below, we shall again write $\psi_0(\br)$ and 
$\psi_1(\br)$, omitting the time variable in order to simplify the 
formulas. Then we find the variational derivatives of the zero-order 
term,
$$
\frac{\dlt H^{(0)}}{\dlt\psi_0^\dgr(\br,t)} = \left ( -\;
\frac{\nabla^2}{2m} + U - \mu_0 - \mu\right ) \psi_0(\br) +
\int \Phi(\br-\br')\psi_0^\dgr(\br')\psi_0(\br')\psi_0(\br) \; d\br' \; ,
$$
of the second-order term,
$$
\frac{\dlt H^{(2)}}{\dlt\psi_0^\dgr(\br,t)} = \int \Phi(\br-\br')
\left [ \psi_1^\dgr(\br')\psi_1(\br')\psi_0(\br) +
\psi_1^\dgr(\br')\psi_0(\br')\psi_1(\br) +
\psi_0^\dgr(\br')\psi_1(\br')\psi_1(\br) \right ] \; d\br' \; ,
$$
and of the third-order term
$$
\frac{\dlt H^{(3)}}{\dlt\psi_0^\dgr(\br,t)} = \int \Phi(\br-\br')
\psi_1^\dgr(\br')\psi_1(\br')\psi_1(\br) \; d\br' \; .
$$
Substituting this into Eq. (65) and taking its coherent average yields
\be
\label{66}
i\; \frac{\prt}{\prt t}\; \eta(\br) = \left ( -\; \frac{\nabla^2}{2m}
+ U - \mu_0 - \mu\right )\eta(\br) + \int \Phi(\br-\br') \left [
|\eta(\br')|^2 \eta(\br) +\hat X(\br,\br')\right ] \; d\br'\; ,
\ee
where
\be
\label{67}
\hat X(\br,\br') \equiv 
\psi_1^\dgr(\br')\psi_1(\br')\eta(\br) +
\psi_1^\dgr(\br')\eta(\br')\psi_1(\br) +
\eta^*(\br')\psi_1(\br')\psi_1(\br) + 
\psi_1^\dgr(\br')\psi_1(\br')\psi_1(\br) \; .
\ee
We may note that Eq. (66) can also be derived from the equation
\be
\label{68}
i\; \frac{\prt}{\prt t}\; \eta(\br,t) = 
\frac{\dlt H[\eta+\psi_1]}{\dlt\eta^*(\br,t)} \; ,
\ee
which follows from taking the coherent average of Eq. (65).

The second equation of motion describes the evolution of the field operator 
of noncondensed particles,
\be
\label{69}
i\; \frac{\prt}{\prt t}\; \psi_1(\br,t) = 
\frac{\dlt H[\psi_0\oplus\psi_1]}{\dlt\psi_1^\dgr(\br,t)} \; .
\ee
With the Hamiltonian (58), we get 
$$
\frac{\dlt H[\psi_0\oplus\psi_1]}{\dlt\psi_1^\dgr(\br,t)} =
\frac{\dlt}{\dlt\psi_1^\dgr(\br,t)}\; \left (
H^{(2)} + H^{(3)} + H^{(4)} \right ) \; .
$$
The corresponding variational derivatives give for the second-order term
$$
\frac{\dlt H^{(2)}}{\dlt\psi_1^\dgr(\br,t)} = \left ( -\;
\frac{\nabla^2}{2m} + U - \mu \right ) \psi_1(\br) +
$$
$$
+ \int \Phi(\br-\br') \left [ 
\psi_0^\dgr(\br')\psi_0(\br')\psi_1(\br) +
\psi_0^\dgr(\br')\psi_1(\br')\psi_0(\br) +
\psi_1^\dgr(\br')\psi_0(\br')\psi_0(\br) \right ] \; d\br' \; ,
$$
for the third-order term,
$$
\frac{\dlt H^{(3)}}{\dlt\psi_1^\dgr(\br,t)} =
 \int \Phi(\br-\br') \left [
\psi_1^\dgr(\br')\psi_1(\br')\psi_0(\br) +
\psi_1^\dgr(\br')\psi_0(\br')\psi_1(\br) +
\psi_0^\dgr(\br')\psi_1(\br')\psi_1(\br) \right ] \; d\br' \; ,
$$
and for the fourth-order term,
$$
\frac{\dlt H^{(4)}}{\dlt\psi_1^\dgr(\br,t)} =
\int \Phi(\br-\br') \psi_1^\dgr(\br')\psi_1(\br')\psi_1(\br)\; d\br' \; .
$$
Taking the coherent average of Eq. (69), we obtain
$$
i\; \frac{\prt}{\prt t}\; \psi_1(\br) = \left ( -\;
\frac{\nabla^2}{2m} + U - \mu\right ) \psi_1(\br) +
$$
\be
\label{70}
+ \int \Phi(\br-\br')\left [ |\eta(\br')|^2 \psi_1(\br) +
\eta^*(\br')\psi_1(\br')\eta(\br) + 
\psi_1^\dgr(\br')\eta(\br')\eta(\br) + \hat X(\br,\br')\right ]\;
d\br' \; .
\ee
The latter equation can also be derived directly from the equation
\be
\label{71}
i\; \frac{\prt}{\prt t} \; \psi_1(\br,t) = 
\frac{\dlt H[\eta+\psi_1]}{\dlt\psi_1^\dgr(\br,t)} \; .
\ee

The final equation for the condensate wave function follows from Eq. (66)
after averaging it over the space $\cF_1$. In doing this, we define for 
the noncondensed particles the normal density matrix
\be
\label{72}
\rho_1(\br,\br') \equiv \; < \psi_1^\dgr(\br')\psi_1(\br)>_{\cF_1} \; ,
\ee
the anomalous density matrix
\be
\label{73}
\sgm_1(\br,\br') \equiv  \; < \psi_1(\br')\psi_1(\br)>_{\cF_1} \; ,
\ee
and the related diagonal densities
\be
\label{74}
\rho_1(\br)\equiv \rho_1(\br,\br)\; , 
\sgm_1(\br)\equiv \sgm_1(\br,\br)\; .
\ee
The total particle density is
\be
\label{75}
\rho(\br) \equiv |\eta(\br)|^2 + \rho_1(\br) \; .
\ee
Averaging Eq. (67) gives
\be
\label{76}
<\hat X(\br,\br')>_{\cF_1} = \rho_1(\br')\eta(\br) 
+\rho_1(\br,\br')\eta(\br') + \sgm_1(\br,\br')\eta^*(\br') + 
<\psi_1^\dgr(\br')\psi_1(\br')\psi_1(\br)>_{\cF_1} \; .
\ee
Then from Eq. (66) we obtain
$$
i\; \frac{\prt}{\prt t}\; \eta(\br) = \left ( -\; \frac{\nabla^2}{2m} +
U - \mu_0 - \mu \right ) \eta(\br) + 
$$
\be
\label{77}
+ \int \Phi(\br-\br') \left [ \rho(\br')\eta(\br) + 
\rho_1(\br,\br')\eta(\br') + \sgm_1(\br,\br')\eta^*(\br') +
<\psi_1^\dgr(\br')\psi_1(\br')\psi_1(\br)>_{\cF_1}\right ]\; d\br' \; .
\ee
This is an exact equation valid for any Bose-condensed system, equilibrium 
or nonequilibrium, uniform or nonuniform.

For an equilibrium system, the coherent field $\eta(\br)$ may be treated 
as not depending on time. Generally speaking, a stationary solution could 
depend on time as $\eta(\br,t)\propto\exp(-i\mu_0't)$. This, however, would
simply lead to a redefinition of the Lagrange multiplier $\mu_0$. Hence, an
equilibrium condensate function can be defined as time independent. In that 
case, Eq. (77) yields
$$
\ep\eta(\br)=\left [ -\; \frac{\nabla^2}{2m} + U(\br)\right ]\eta(\br) +
$$
\be
\label{78}
+ \int \Phi(\br-\br') \left [ \rho(\br')\eta(\br) +
\rho_1(\br,\br')\eta(\br') + \sgm_1(\br,\br')\eta^*(\br') +
<\psi_1^\dgr(\br')\psi_1(\br')\psi_1(\br)>_{\cF_1} \right ]\; d\br' \; ,
\ee
where the notation
\be
\label{79}
\ep \equiv \mu_0 + \mu
\ee
is introduced, playing the role of the condensate energy per particle. 
Equation (78) is a generalized eigenproblem, which can possess the whole 
spectrum of the energies $\ep$ and the related eigenfunctions $\eta(\br)$. 
In the case of a confining potential $U(\br)$, representing an atomic trap, 
the spectrum $\ep$ is discrete, with the solutions $\eta(\br)$ representing 
topological coherent modes [38--41]. In equilibrium, the lowest-energy 
mode corresponds to the usual condensate. But in a nonequilibrium system, 
nonground-state condensates can be generated [38--41].

In a uniform system, where $U(\br)=0$, we have $\eta(\br)=\eta$, 
$\rho(\br)=\rho$, with $\rho_1(\br,\br')$ and $\sgm_1(\br,\br')$ 
depending on the difference $\br-\br'$. Then Eq. (78) reduces to
\be
\label{80}
\ep\eta = \rho\Phi_0\eta +
\int \Phi(\br) \left [ \rho_1(\br,0)\eta + \sgm_1(\br,0)\eta^* +
<\psi_1^\dgr(0)\psi_1(0)\psi_1(\br)>_{\cF_1} \right ] \; d\br \; .
\ee
Recall that Eqs. (77), (78), and (80) for the corresponding systems are
exact, with no approximations involved.

\section{Dilute gas}

To illustrate the above equations for some particular cases, let us resort 
to the dilute-gas approximation, when the inequality
\be
\label{81}
|\rho a_s^3|\ll 1
\ee
is valid, where $\rho\equiv N/V$ is the average density and $a_s$ is the 
$s$-wave scattering length. Then the interaction potential can be modelled 
by the contact form
\be
\label{82}
\Phi(\br) = \Phi_0\dlt(\br) \; ,
\ee
with
$$
\Phi_0 \equiv \int \Phi(\br)\; d\br = 4\pi\; \frac{a_s}{m} \; .
$$
The condensate-function equation (77) 
becomes
$$
i\; \frac{\prt}{\prt t}\; \eta(\br) = \left ( - \; \frac{\nabla^2}{2m}
+ U - \ep\right )\eta(\br) +
$$
\be
\label{83}
+ \Phi_0\left [ \rho(\br)\eta(\br) + \rho_1(\br)\eta(\br) +
\sgm_1(\br)\eta^*(\br) + <\psi_1^\dgr(\br)\psi_1(\br)\psi_1(\br)>
\right ] \; .
\ee
The equation of motion (70) for the field operator of noncondensed particles
takes the form
\be
\label{84}
i\; \frac{\prt}{\prt t}\; \psi_1(\br) = \left ( - \; \frac{\nabla^2}{2m}
+ U - \mu \right )\psi_1(\br) +
\Phi_0\left [ 2|\eta(\br)|^2\psi_1(\br) + \eta^2(\br)\psi_1^\dgr(\br) +
\hat X(\br,\br) \right ] \; ,
\ee
where 
$$ 
\hat X(\br,\br)=2\psi_1^\dgr(\br)\psi_1(\br)\eta(\br) +
\eta^*(\br)\psi_1(\br)\psi_1(\br) + 
\psi_1^\dgr(\br)\psi_1(\br)\psi_1(\br) \; .
$$

In Eq. (83) and in what follows, we shall not specify explicitly the Fock 
spaces over which the averages are to be taken. This makes it possible to 
simplify the notation and cannot lead to confusion if, by definition, we 
accept the rule that each average of operators is calculated in that Fock 
space, where these operators are defined.

Further simplification can be done by employing the Hartree-Fock-Bogolubov
(HFB) approximation. Then
$$
<\psi_1^\dgr(\br)\psi_1(\br)\psi_1(\br)>\; = \; 0 \; .
$$
And the condensate-function equation (83) reduces to
\be
\label{85}
i\; \frac{\prt}{\prt t}\; \eta(\br) = \left ( - \; \frac{\nabla^2}{2m}
+ U - \ep \right )\eta(\br) + \Phi_0\left [ \rho(\br)\eta(\br) + 
\rho_1(\br)\eta(\br) + \sgm_1(\br)\eta^*(\br) \right ] \; .
\ee
If one considers the case, when all particles are condensed, that is,
$\rho(\br)=|\eta(\br)|^2$ and $\rho_1(\br)=\sgm_1(\br)=0$, then Eq. (85)
simplifies to the equation
$$
i\; \frac{\prt}{\prt t}\; \eta(\br) = \left [ - \; \frac{\nabla^2}{2m}
+ U - \ep +\Phi_0|\eta(\br)|^2\right ] \eta(\br) \; .
$$
The latter is what is generally known as the nonlinear Schr\"odinger
equation and in condensed-matter literature as the Gross-Pitaevskii 
equation.

In the HFB approximation, the terms entering $\hat X(\br,\br)$ linearize 
as
$$
\psi_1^\dgr(\br)\psi_1(\br) = \psi_1^\dgr(\br)<\psi_1(\br)> +
< \psi_1^\dgr(\br)>\psi_1(\br) -  <\psi_1^\dgr(\br)><\psi_1(\br)>\; =
\; 0
$$
and
$$
\psi_1^\dgr(\br)\psi_1(\br)\psi_1(\br) = 2\rho_1(\br)\psi_1(\br)
+\sgm_1(\br)\psi_1^\dgr(\br) \; ,
$$
where the conservation condition (39) is taken into account. The linearized
Eq. (84) is
\be
\label{86}
i\; \frac{\prt}{\prt t}\; \psi_1(\br) = \left ( - \; \frac{\nabla^2}{2m}
+ U - \mu \right ) \psi_1(\br) + \Phi_0\left [ 2\rho(\br)\psi_1(\br) +
\sgm(\br)\psi_1^\dgr(\br) \right ] \; ,
\ee
where
\be
\label{87}
\sgm(\br) \equiv \eta^2(\br) +\sgm_1(\br) \; .
\ee

For an equilibrium system, the condensate-function equation (85) becomes
\be
\label{88}
\ep\eta =  \left [ - \; \frac{\nabla^2}{2m}+ U(\br)\right ]\eta(\br) +
\Phi_0 \left [ \rho(\br)\eta(\br) + \rho_1(\br)\eta(\br) +
\sgm_1(\br)\eta^*(\br) \right ] \; .
\ee
In the case of a uniform system, when $U(\br)=0$, the condensate amplitude 
$\eta$ can be chosen real. From Eq. (88), taking into consideration that
$\rho_1(\br)=\rho_1$ and $\sgm_1(\br)=\sgm_1$, we have
\be
\label{89}
\ep = ( \rho+\rho_1+\sgm_1) \Phi_0 \; .
\ee

Let us consider in more detail an equilibrium uniform system. The field 
operator of noncondensed particles can be expanded over the basis of 
plane waves
$$
\vp_k(\br) = \frac{1}{\sqrt{V}}\; e^{i\bk\cdot\br}
$$
as
\be
\label{90}
\psi_1(\br) = \sum_{k\neq 0} a_k \vp_k(\br) \; .
\ee
Because of the global symmetry of the Hamiltonian with respect to the 
transformation
\be
\label{91}
\eta(\br) \longrightarrow \eta(\br) e^{i\al}\; , \qquad
\psi_1(\br) \longrightarrow \psi_1(\br)  e^{i\al}\; ,
\ee
in which $\al$ is a real number, the condensate amplitude $\eta(\br)=\eta$
can be taken real,
\be
\label{92}
\eta =\eta^* =\sqrt{\rho_0} \; , \qquad \rho_0\equiv \frac{N_0}{V} \; .
\ee

The field operators $a_k$ and $a_k^\dgr$ in the momentum representation are 
defined on the Fock space $\cF(a_k)$ generated by $a_k^\dgr$. An important
role is played by the following averages. The momentum distribution of
particles
\be
\label{93}
n_k \equiv \; < a_k^\dgr a_k>
\ee
is called the normal average, as opposed to
\be
\label{94}
\sgm_k \equiv \; < a_k a_{-k}>\; ,
\ee
termed the anomalous average. The latter gives the amplitude of the process, 
when two particles are annihilated from the thermal cloud of noncondensed
particles. Respectively, $\sgm_k^*=<a_k^\dgr a_{-k}^\dgr>$ is the amplitude
for the process of creation of two noncondensed particles. With $\sgm_k$ 
being the amplitude of these pair processes of creation and annihilation, 
the absolute value $|\sgm_k|$ describes the density of binary correlated 
particles.

The averages (93) and (94) define the density of noncondensed particles,
\be
\label{95}
\rho_1 \equiv \; <\psi_1^\dgr(\br)\psi_1(\br)>\; = \;
\frac{1}{V}\; \sum_{k\neq 0} n_k \; ,
\ee
and the amplitude of binary correlated particles,
\be
\label{96}
\sgm_1\equiv \; <\psi_1(\br)\psi_1(\br)>\; = \; 
\frac{1}{V}\; \sum_{k\neq 0} \sgm_k \; .
\ee
The total number of noncondensed particles is
\be
\label{97}
N_1  = \sum_{k\neq 0} n_k
\ee
and the number of binary correlated particles is
\be
\label{98}
B_1  = \left | \sum_{k\neq 0} \sgm_k\right | \; .
\ee
So that the number of correlated pairs is $B_1/2$.

Substituting the field-operator expansion (90) into the Hamiltonian (58),
we take the coherent average, as a result of which the condensate operators
$\psi_0(\br)$ are replaced by the coherent field $\eta(\br)$. For the 
considered uniform system, we find the zero-order term
\be
\label{99}
H^{(0)} = \left ( \frac{1}{2}\; \rho_0\Phi_0 - \ep\right ) N_0 \; .
\ee
The first-order term $H^{(1)}$ is exactly zero. The second-order term is
\be
\label{100}
H^{(2)} = \sum_{k\neq 0} \left ( \frac{k^2}{2m} + 2\rho_0\Phi_0 -
\mu\right ) a_k^\dgr a_k + \frac{1}{2}\; \rho_0 \Phi_0 \;
\sum_{k\neq 0} \left ( a_k^\dgr a_{-k}^\dgr + a_{-k}  a_k\right ) \; .
\ee
The sum of Eqs. (99) and (100) corresponds to the Hamiltonian in the 
Bogolubov approximation [7--10]. For the higher-order terms, we obtain 
the following. The third-order term is
\be
\label{101}
H^{(3)}  =\sqrt{\frac{\rho_0}{V}}\; \Phi_0 {\sum_{k,p}}' \left (
a_k^\dgr a_{k+p}a_{-p} +  a_{-p}^\dgr a_{k+p}^\dgr a_k \right ) \; ,
\ee
where the prime at the summation sign indicates that no operator in the
summation corresponds to $a_0$, so that here
$$
\bk \neq 0 \; , \qquad {\bf p}\neq  0 \; , \qquad
\bk +  {\bf p}\neq  0 \; .
$$
For the fourth-order term, we get
\be
\label{102}
H^{(4)} = \frac{\Phi_0}{2V} \; \sum_k \; {\sum_{p,q}}' 
a_p^\dgr a_q^\dgr a_{k+p} a_{q-k} \; ,
\ee
where the prime again means the same as earlier, that is, here it implies
that
$$
{\bf p}\neq  0 \; , \qquad {\bf q}\neq  0 \; , \qquad
\bk + {\bf p}\neq  0 \; , \qquad {\bf q} -\bk \neq 0 \; .
$$

The higher-order terms (101) and (102) can be treated in the HFB 
approximation.  In so doing, we keep in mind that, as a consequence of 
the conservation condition (39), we have
\be
\label{103}
<a_k>\; = \; 0 \qquad (\bk \neq 0) \; .
\ee
Therefore Eq. (101) vanishes,
\be
\label{104}
H^{(3)} = 0\; .
\ee
And Eq. (102) reduces to
\be
\label{105}
H^{(4)} = \Phi_0\; \sum_{k\neq 0} \left ( 2\rho_1 a_k^\dgr a_k + 
\frac{1}{2}\; \sgm_1 a_k^\dgr a_{-k}^\dgr + \frac{1}{2}\; \sgm_1^* 
a_{-k} a_k \right )\; - \; \frac{1}{2} \left ( 2\rho_1^2 + |\sgm_1|^2
\right ) \Phi_0 V \; .
\ee

In this way, instead of the Hamiltonian (58), we have in the HFB 
approximation
\be
\label{106}
H_{HFB} =  H^{(0)} + H^{(2)} + H^{(4)} \; ,
\ee
where $H^{(0)}$ is given by Eq. (99); $H^{(2)}$, by Eq. (100); and 
$H^{(4)}$, by Eq. (105). To write down the Hamiltonian (106) in a compact 
form, let us introduce the notation
\be
\label{107}
\om_k \equiv \frac{k^2}{2m} + 2\rho\Phi_0 - \mu 
\ee
and, assuming that $\sgm_1$ is real, we define
\be
\label{108}
\Dlt \equiv (\rho_0 +\sgm_1) \Phi_0 \; .
\ee
Then Hamiltonian (106) writes as
\be
\label{109}
H_{HFB} = E_{HFB} + \sum_{k\neq 0} \om_k a_k^\dgr a_k + 
\frac{1}{2}\; \Dlt \; \sum_{k\neq 0} \left ( a_k^\dgr a_{-k}^\dgr +
a_{-k} a_k \right ) \; ,
\ee
with the HFB nonoperator term
$$
E_{HFB} \equiv H^{(0)} -\; \frac{1}{2}\; \left ( 2\rho_1^2 + \sgm_1^2
\right ) \Phi_0 V \; .
$$

Invoking the Bogolubov canonical transformation (21), discussed in 
Section 3, Hamiltonian (109) is diagonalized to the Bogolubov form
\be
\label{110}
H_B = E_0 + \sum_{k\neq 0} \ep_k b_k^\dgr b_k \; ,
\ee
with the ground-state energy
$$
E_0 = E_{HFB} + \frac{1}{2} \; \sum_{k\neq 0} (\ep_k - \om_k)
$$
and with the spectrum 
\be
\label{111}
\ep_k =\sqrt{\om_k^2-\Dlt^2} \; .
\ee

Separation of Bose-Einstein condensate by means of the Bogolubov shift 
is meaningful only when the momentum distribution of particles displays 
a singularity at the point $k=0$, which happens if the quasiparticle 
spectrum, being positive, touches zero at zero momentum [42]. That is,
the necessary condition for the occurrence of Bose-Einstein condensation
is
\be
\label{112}
\lim_{k\ra 0} \ep_k = 0 \; , \qquad \ep_k \geq 0 \; .
\ee
This defines the chemical potential
$$
\mu = 2\rho\Phi_0 - \Dlt \; ,
$$
which, with Eq. (108), becomes
\be
\label{113}
\mu = (\rho+\rho_1-\sgm_1) \Phi_0 \; .
\ee
Combining this with Eqs. (79) and (89), we get
\be
\label{114}
\mu_0 = \ep -\mu = 2\sgm_1\Phi_0 \; .
\ee
Substituting Eq. (113) into Eq. (107), we have
$$
\om_k = \frac{k^2}{2m} + (\rho_0 +\sgm_1)\Phi_0 \; ,
$$
which, taking into consideration Eqs. (108) and (111), can be written as
$$
\om_k =\frac{k^2}{2m} + \Dlt =\sqrt{\ep_k^2 +\Dlt^2} \; .
$$
The quasiparticle spectrum (111) can be represented in the standard form 
of the Bogolubov spectrum 
\be
\label{115}
\ep_k =\sqrt{(ck)^2 +\left ( \frac{k^2}{2m}\right )^2 } \; ,
\ee
with the sound velocity given by the equation
\be
\label{116}
c \equiv \sqrt{\frac{\Dlt}{m} } \qquad \left (\Dlt=mc^2\right ) \; .
\ee

As is evident, spectrum (115) is gapless, in agreement with the theorems
by Bogolubov [9,10] and Hugenholtz and Pines [14]. At the same time, we 
have encountered no contradiction with the equations of motion. Recall 
that the latter were derived from one given Hamiltonian, because of which 
these equations automatically preserve all conservation laws.

The diagonal Hamiltonian (110) makes it easy to calculate different averages 
in the space $\cF(b_k)$. Thus, the momentum distribution of quasiparticles is
\be
\label{117}
\pi_k \equiv \; <b_k^\dgr b_k> \; = \; \left ( e^{\bt\ep_k} -1
\right )^{-1} \; .
\ee
Taking into account that
$$
1 + 2\pi_k = {\rm coth}\left ( \frac{\bt\ep_k}{2}\right ) \; ,
$$
we find the momentum distribution of particles (93) as
\be
\label{118}
n_k = \frac{\sqrt{\ep_k^2+\Dlt^2}}{2\ep_k}\; {\rm coth}\left (
\frac{\bt\ep_k}{2}\right ) - \; \frac{1}{2}
\ee
and the anomalous average (94) as
\be
\label{119}
\sgm_k = -\; \frac{\Dlt}{2\ep_k}\;  {\rm coth}\left (
\frac{\bt\ep_k}{2}\right ) \; .
\ee

For the density of noncondensed particles (95), we obtain
\be
\label{120}
\rho_1 = \frac{1}{2} \; \int \left [ 
\frac{ \sqrt{\ep_k^2+\Dlt^2}}{\ep_k} \;  
{\rm coth}\left ( \frac{\bt\ep_k}{2}\right ) -1 \right ] \;
\frac{d\bk}{(2\pi)^3} \; ,
\ee
and for the anomalous amplitude (98), we find
\be
\label{121}
\sgm_1 = -(\rho_0+\sgm_1) \int \frac{\Phi_0}{2\ep_k}\; {\rm coth}\left ( 
\frac{\bt\ep_k}{2}\right ) \;\frac{d\bk}{(2\pi)^3} \; .
\ee
In agreement with the earlier assumption, the amplitude $\sgm_1$ can 
be chosen to be real. However, the integral in Eq. (121) displays an 
ultraviolet divergence caused by the dilute-gas approximation with the 
contact potential (82). This artificial divergence can be easily removed 
by accepting a more realistic interaction potential $\Phi(\br)$ possessing
a nontrivial momentum dependence of its Fourier transform $\Phi_k$. One
often takes the potential $\Phi(\br)$ in the Gaussian form [43--45], for
which $\Phi_k$ exponentially decreases as $k\ra\infty$. Replacing in Eq.
(108) $\Phi_0$ by $\Phi_k$, we get
\be
\label{122}
\Dlt_k \equiv (\rho_0 +\sgm_1) \Phi_k \; .
\ee
Now, the integral in Eq. (121) becomes
\be
\label{123}
\al \equiv \int \frac{\Phi_k}{2\ep_k}\; {\rm coth}\left (
\frac{\bt\ep_k}{2}\right )\; \frac{d\bk}{(2\pi)^3} \; .
\ee
The latter is convergent as soon as $\Phi_k$ tends to zero faster than 
$1/k^2$.

Another possibility of removing the ultraviolet divergence could be by 
subtracting the divergent term in Eq. (121), as is often done when 
considering hard-core particles. In such a manner, integral (123) could 
be defined as
$$
\al = \int \left ( \frac{\Phi_0}{2\ep_k} \; - \; \frac{m\Phi_0}{k^2} 
\right ) {\rm coth}\left ( \frac{\bt\ep_k}{2}\right )\; 
\frac{d\bk}{(2\pi)^3} \; .
$$
Note, however, that the latter expression becomes negative, since 
$\ep_k>k^2/2m$,
while Eq. (123) is positive.

Using notation (123), we obtain, instead of Eq. (121), the relation
$$
\sgm_1 = -(\rho_0 + \sgm_1) \al \; .
$$
Solving the latter, we find
\be
\label{124}
\sgm_1  = -\; \frac{\rho_0\al}{1+\al} \; .
\ee
Because integral (123) is positive, the anomalous amplitude (124) is 
negative. From Eq. (122), we have
\be
\label{125}
\Dlt_k = \frac{\rho_0\Phi_k}{1+\al} \; ,
\ee
which is positive for all $k$. As is seen from Eqs. (124) and (125), in
vicinity of the condensation temperature $T_c$, defined by the equation 
$\rho_1(T_c)=\rho$, one has $\rho_0\ra 0$ as $T\ra T_c$. Hence both $\sgm_1$ 
and $\Dlt_k$ tend to zero when $T\ra T_c$. But for lower temperatures 
$T\ll T_c$, where $\rho_0$ can be close to $\rho$, the anomalous amplitude 
$\sgm_1$ becomes of the order of $\rho_1$ and can even be much larger 
than the latter. Therefore the anomalous averages cannot be neglected at
low temperatures.

It is necessary to stress the difference between the theory, developed 
in this paper, and the standard approach to Bose-condensed systems. The 
principal difference is that in the present theory two Lagrange multipliers, 
$\mu_0$ and $\mu$, are introduced in Eq. (57) in order to preserve two 
normalization conditions (45) and (55). While in the standard theory, 
solely the chemical potential $\mu$ is introduced. because of this, the
HFB approximation in the standard theory acquires an unphysical gap in 
the spectrum of collective excitations. If one removes the gap by some
additional tricks, the theory becomes nonconserving. Thus, the standard 
approach always suffers from one of the defficiencies, being either 
gapeful or nonconserving [15].

Sometimes, to remove the gap in the collective spectrum, one sets to zero 
the anomalous average (119), ascribing this trick to Popov. First of all, 
such a trick is principally incorrect at low temperatures, when the anomalous 
averages become of the order or larger than the normal averages [46]. And, 
moreover, as is easy to infer from the original works by Popov [16--19], he 
has never suggested to use such an incorrect trick at low temperatures.

In the theory, presented in this paper, there is no need to invoke any 
unjustified tricks. All averages, normal as well as anomalous, are treated 
on an equal footing. And the theory always remains both gapless and 
conserving.

In conclusion to this section, it is useful to analyse the stability of 
matter in the HFB approximation. As is well known [47], the ideal uniform 
Bose-condensed gas is unstable. This instability stems from the divergent
compressibility, which is directly related to anomalous particle 
fluctuations [5].

There have been published quite a number of papers claiming that particle 
fluctuations in the interacting Bose gas remain anomalous, being proportional
to $N^{4/3}$ in the Bogolubov approximation. This type of anomalous behaviour 
does not depend on the ensemble employed, being the same in the grand 
canonical and canonical Gibbs ensembles, as well as in the microcanonical
ensemble [48]. If such anomalous fluctuations would really be present, 
this would mean the instability of the system [5,49,50].

However, as has been thoroughly explained in Ref. [5,49,50], there are no 
anomalous fluctuations in the Bogolubov approximation. Particle fluctuations
in the Bogolubov theory are normal, being proportional to $N$. The anomalies
in the fluctuations arise solely to the unjustified usage of an approximation
outside of the region of its validity [49,50].

Let us consider particle fluctuations in the HFB approximation. These 
fluctuations are characterized by the dispersion
\be
\label{126}
\Dlt^2(\hat N) \equiv \; < \hat N^2> - < \hat N>^2 \; .
\ee
A convenient expression for this dispersion is given [5,49] by the 
equation
\be
\label{127}
\Dlt^2(\hat N) = N \left \{ 1 + \rho \int
[g(\br)-1]\; d\br\right \} \; ,
\ee
in which the pair correlation function
\be
\label{128}
g(\br) \equiv \frac{1}{\rho^2} < \tilde\psi^\dgr(\br) \tilde\psi^\dgr(0)
\tilde\psi(0) \tilde\psi(\br)>
\ee
is defined through the field operators (42).

Calculating function (128), we should keep in mind that the HFB 
approximation is equivalent to an effective mean-field theory, with 
a quadratic Hamiltonian in terms of the field operators $\psi_1(\br)$. 
Therefore, substituting the representation (42) into Eq. (128), we should 
retain there only the terms up to the second order with respect to 
$\psi_1(\br)$. Then we have
$$
g(\br) =  1 + \frac{2\rho_0}{\rho^2}\; [\rho_1(\br,0) +\sgm_1(\br,0)] \; .
$$
Because of the same reason, we should set here $\rho_0/\rho\ra 1$, since
$\rho_0\equiv\rho-\rho_1$. Thus, we get
\be
\label{129}
g(\br) =  1 + \frac{2}{\rho} \; \int (n_k+\sgm_k)\;
e^{i\bk\cdot\br}\; \frac{d\bk}{(2\pi)^3} \; .
\ee
Using the expressions for $n_k$ and $\sgm_k$, obtained above, we find
\be
\label{130}
\Dlt^2(\hat N) = \frac{TN}{mc^2} \; .
\ee
That is, particle fluctuations in the HFB approximation are normal, in 
the same way as in the Bogolubov approximation [5,49,50].

\section{Nonuniform matter}

The consideration of the previous Section 6 can be generalized to the 
case of nonuniform Bose systems with arbitrary interaction potentials 
$\Phi(\br)$. As is shown in Section 5, the final equations of motion can 
be derived by projecting the Hamiltonian (58) on the space $\cF_1$. This
implies that the coherent average of Eq. (58) is taken in the space $\cF_0$, 
which results in the replacement of the condensate field operators 
$\psi_0(\br)$ by the condensate wave function $\eta(\br)$. Keeping this
projection in mind, we shall work now directly in the space $\cF_1$.

For concreteness, we shall employ the HFB approximation, which for the 
fourth-order product of the noncondensed-particle field operators writes 
as
$$
\psi_1^\dgr(\br)\psi_1^\dgr(\br') \psi_1(\br') \psi_1(\br)  =
\rho_1(\br)\psi_1^\dgr(\br') \psi_1(\br') + 
\rho_1(\br')\psi_1^\dgr(\br) \psi_1(\br) +
$$
$$
+ \rho_1(\br',\br)\psi_1^\dgr(\br') \psi_1(\br) +
\rho_1(\br,\br')\psi_1^\dgr(\br) \psi_1(\br') +
$$
\be
\label{131}
+ \sgm_1(\br,\br')\psi_1^\dgr(\br) \psi_1^\dgr(\br')+
\sgm_1^*(\br',\br)\psi_1(\br') \psi_1(\br) -
\rho_1(\br)\rho_1(\br') -|\rho_1(\br,\br')|^2 -
|\sgm_1(\br,\br')|^2 \; ,
\ee
where the notation from Eqs. (72) to (74) is used. Then, instead of Eq. 
(109), we come to the HFB Hamiltonian for nonuniform matter
$$
H_{HFB} = E_{HFB} + \int \psi_1^\dgr(\br) \left ( - \;
\frac{\nabla^2}{2m} + U - \mu\right )\psi_1(\br)\; d\br +
\int \Phi(\br-\br') \left [ \rho_1(\br')\psi_1^\dgr(\br) \psi_1(\br) +
\right.
$$
\be
\label{132}
\left.
+ \rho(\br',\br) \psi_1^\dgr(\br') \psi_1(\br) + \frac{1}{2}\;
\sgm(\br,\br') \psi_1^\dgr(\br') \psi_1^\dgr(\br) + \frac{1}{2}\;
\sgm^*(\br,\br') \psi_1(\br') \psi_1(\br) \right ]\; d\br d\br' \; ,
\ee
in which the notation
\be
\label{133}
\rho(\br,\br') \equiv \eta(\br)\eta^*(\br') +\rho_1(\br,\br')
\ee
and
\be
\label{134}
\sgm(\br,\br') \equiv \eta(\br)\eta(\br') +\sgm_1(\br,\br')
\ee
is introduced and where
\be
\label{135}
E_{HFB} \equiv H^{(0)} - \; \frac{1}{2}\; \int \Phi(\br-\br')\left [
\rho_1(\br)\rho_1(\br') + |\rho_1(\br,\br')|^2 + |\sgm_1(\br,\br')|^2
\right ] \; d\br d\br' \; ,
\ee
with $H^{(0)}$ given in Eq. (99).

The quadratic form (132) can be diagonalized by means of the general 
canonical transformations [20]. For this purpose, we may expand the 
operators $\psi_1(\br)$ and $\psi_1^\dgr(\br)$ as
$$
\psi_1(\br) = \sum_k \left [ u_k(\br)b_k +
v_k^*(\br) b_k^\dgr\right ]\; ,
$$
\be
\label{136}
\psi_1^\dgr(\br) = \sum_k \left [ u_k^*(\br)b_k^\dgr + 
v_k(\br) b_k\right ]\; ,
\ee
where $k$ is a set of quantum numbers. The transformations, inverse to 
Eq. (136), are
$$
b_k = \int \left [ u_k^*(\br)\psi_1(\br) -
v_k^*(\br)\psi_1^\dgr(\br)\right ]\; d\br \; ,
$$
\be
\label{137}
b_k^\dgr = \int \left [ u_k(\br)\psi_1^\dgr(\br) -
v_k(\br)\psi_1(\br)\right ]\; d\br \; .
\ee
The operators $\psi_1(\br)$ and $b_k$ are assumed to satisfy the Bose 
commutation relations, which imposes on the coefficient functions $u_k(\br)$ 
and $v_k(\br)$ the following restrictions:
$$
\sum_k \left [ u_k(\br) v_k^*(\br') - v_k^*(\br) u_k(\br')\right ] = 0 \; ,
$$
$$
\sum_k \left [ u_k(\br) u_k^*(\br') - v_k^*(\br) v_k(\br')\right ] = 
\dlt(\br-\br') \; ,
$$
$$
\int \left [ u_k(\br) v_p(\br) - v_k(\br) u_p(\br)\right ]\; d\br = 0 \; ,
$$
\be
\label{138}
\int \left [ u_k^*(\br) u_p(\br) - v_k^*(\br) v_p(\br)\right ]\; d\br =  
\dlt_{kp} \; .
\ee
Let us define
\be
\label{139}
\om(\br,\br') \equiv \left [ -\; \frac{\nabla^2}{2m} + U(\br) - \mu +
\int \Phi(\br-\br')\rho(\br')\; d\br'\right ]\dlt(\br-\br') +
\Phi(\br-\br')\rho(\br,\br')
\ee
and
\be
\label{140}
\Dlt(\br,\br') \equiv \Phi(\br-\br')\sgm(\br,\br') \; .
\ee
Then the diagonalization condition can be represented as a system of 
equations
$$
\int \left [ \om(\br,\br') u_k(\br') +\Dlt(\br,\br') v_k(\br')
\right ]\; d\br' = \ep_k u_k(\br) \; ,
$$
\be
\label{141}
\int \left [ \om^*(\br,\br') v_k(\br') +\Dlt^*(\br,\br') u_k(\br')
\right ]\; d\br' = - \ep_k v_k(\br) \; ,
\ee
which is a variant of the Bogolubov - de Gennes equations. The HFB 
Hamiltonian (132) acquires the diagonal form
\be
\label{142}
H_B = E_0 + \sum_k \ep_k b_k^\dgr b_k \; ,
\ee
which is analogous to Eq. (110), with the difference that here $k$ is not 
momentum but a set of quantum numbers labelling the functions $u_k(\br)$
and $v_k(\br)$. And the ground-state energy is
\be
\label{143}
E_0 = E_{HFB} - \sum_k \ep_k \int |v_k(\br)|^2\; d\br \; .
\ee

Calculating the averages in the space $\cF(b_k)$, with the Hamiltonian 
(142), we find the normal density matrix
\be
\label{144}
\rho_1(\br,\br') = \sum_k \left [ \pi_k u_k(\br) u_k^*(\br') +
(1+\pi_k)v_k^*(\br) v_k(\br') \right ]
\ee
and the anomalous density matrix
\be
\label{145}
\sgm_1(\br,\br') = \sum_k \left [ \pi_k u_k(\br) v_k^*(\br') +
(1+\pi_k)v_k^*(\br) u_k(\br') \right ] \; ,
\ee
defined in Eqs. (72) and (73), with $\pi_k$ from Eq. (117). The corresponding
diagonal elements (74) give the density of noncondensed particles
\be
\label{146}
\rho_1(\br)  =\sum_k \left [ \pi_k|u_k(\br)|^2 +
(1+\pi_k)|v_k(\br)|^2 \right ]
\ee
and the anomalous amplitude
\be
\label{147}
\sgm_1(\br) = \sum_k(1+2\pi_k)u_k(\br) v_k^*(\br) \; .
\ee
These general equations are valid for an arbitrary external potential 
$U(\br)$ and for any pair interaction potential $\Phi(\br)$.

One can always return to the uniform system by taking
\be
\label{148}
u_k(\br) = u_k\vp_k(\br)\; , \qquad v_k(\br) = v_k\vp_k(\br) \; ,
\ee
where $\vp_k(\br)$ is a plane wave, By expanding in the plane waves the 
kernel (139) as
\be
\label{149}
\om(\br,\br') = \sum_k \om_k\vp_k(\br)\vp_k^*(\br')
\ee
and Eq. (140) as
\be
\label{150}
\Dlt(\br,\br') = \sum_k \Dlt_k\vp_k(\br)\vp_k^*(\br') \; ,
\ee
we obtain $\om_k$ and $\Dlt_k$, which are equivalent to Eqs. (107) and 
(108), respectively.

In this way, any system with Bose-Einstein condensate can be described by 
a self-consistent theory, which is both {\it conserving} and {\it gapless}. 
The basis for this theory is an accurate usage of the nonequivalent operator 
representations, associated with canonical commutation relations, and the 
derivation of equations of motion taking into consideration all conditions 
characterizing the Bose-condensed system. In particular, taking into account 
two normalization conditions (45) and (55) requires to introduce two 
Lagrange multipliers $\mu_0$ and $\mu$.

\vskip 5mm

{\bf Acknowledgement}

\vskip 2mm

I appreciate useful discussions with H. Kleinert. I am grateful to the 
German Research Foundation for the Mercator Professorship.

\end{document}